\documentclass{aa} 
\usepackage{psfig}
\begin{document}
\def\lya{Ly$\alpha$}     %
\def\lyb{Ly$\beta$}      %
\def\lyc{Ly$\gamma$}     %
\def\lye{Ly$\epsilon$}   %
\def\Wr{W_\mathrm{r}}   %
\def\Wo{W_\mathrm{obs}} %
\def\mr{m_\mathrm{R}}   %
\def\Mr{M_\mathrm{R}}   %
\def\mb{m_\mathrm{B}}   %
\def\Mb{M_\mathrm{B}}   %
\def\mi{m_\mathrm{I}}   %
\def\Mi{M_\mathrm{I}}   %
\def\mq{m_{s\mathrm 450}}%
\def\ms{m_\mathrm{702}} %
\def\mh{m_\mathrm{814}} %
\def\kms{~km~s$^{-1}$}   %
\def\cm2{~cm$^{-2}$}     %
\def\za{z_\mathrm{a}}   %
\def\zd{z_\mathrm{d}}   %
\def\ze{z_\mathrm{e}}   %
\def\zg{z_\mathrm{g}}   %
\def\h50{h_{50}^{-1}}    %
\def\hi{H\,{\sc i}}      %
\def\ci{C\,{\sc i}}      %
\def\nhi{N(\mbox{H\,{\sc i}})}%
\def\nhii{N(\mbox{H\,{\sc ii}})}%
\def\nci{N(\mbox{C\,{\sc i}})}%
\def\nsiii{N(\mbox{Si\,{\sc ii}})}%
\def\nfeii{N(\mbox{Fe\,{\sc ii}})}%
\def\nsiii{N(\mbox{Si\,{\sc ii}})}%
\def\nmnii{N(\mbox{Mn\,{\sc ii}})}%
\def\nniii{N(\mbox{Ni\,{\sc ii}})}%
\def\nznii{N(\mbox{Zn\,{\sc ii}})}%
\def\lnhi{\log(N(\mbox{H\,{\sc i}}))}%
\def\lnfeii{\log(N(\mbox{Fe\,{\sc ii}}))}%
\def\lnmgii{\log(N(\mbox{Mg\,{\sc ii}}))}%
\def\lnsiii{\log(N(\mbox{Si\,{\sc ii}}))}%
\def\lnniii{\log(N(\mbox{Ni\,{\sc ii}}))}%
\def\lncrii{\log(N(\mbox{Cr\,{\sc ii}}))}%
\def\lncaii{\log(N(\mbox{Ca\,{\sc ii}}))}%
\def\lnznii{\log(N(\mbox{Zn\,{\sc ii}}))}%
\def\lnmnii{\log(N(\mbox{Mn\,{\sc ii}}))}%
\def\alii{Al\,{\sc ii}}  %
\def\aliii{Al\,{\sc iii}}  %
\def\caii{Ca\,{\sc ii}}  %
\def\caiii{Ca\,{\sc iii}}%
\def\ci{C\,{\sc i}}    %
\def\cii{C\,{\sc ii}}    %
\def\ciii{C\,{\sc iii}}  %
\def\civ{C\,{\sc iv}}    %
\def\crii{Cr\,{\sc ii}}  %
\def\feii{Fe\,{\sc ii}}  %
\def\siii{Si\,{\sc ii}}  %
\def\siiii{Si\,{\sc iii}}%
\def\siiv{Si\,{\sc iv}}  %
\def\mgi{Mg\,{\sc i}}    %
\def\mgii{Mg\,{\sc ii}}  %
\def\oi{O\,{\sc i}}      %
\def\ni{N\,{\sc i}}      %
\def\ovi{O\,{\sc vi}}    %
\def\nv{N\,{\sc v}}      %
\def\znii{Zn\,{\sc ii}}  %
\def\niii{Ni\,{\sc ii}}  %
\def\crii{Cr\,{\sc ii}}  %
\def\mnii{Mn\,{\sc ii}}  %
\thesaurus {03(11.17.1; 11.08.1; 11.09.4)} 
\title{Discovery of a $z=0.808$ damped \lya\ system candidate in a UV selected 
quasar spectrum} 
\author{V. Le Brun, M. Viton, B. Milliard}   
\institute{
IGRAP/Laboratoire d'Astronomie Spatiale du C.N.R.S., B.P. 8, F-13376 
Marseille, France ; vlebrun, mviton, bmilliard@astrsp-mrs.fr
} 
\offprints{V. Le Brun}
\date {Received 8 April 1998 ; accepted 6 May 1998}
\maketitle
\markboth{}{}
\maketitle        
\begin{abstract} 
	We present the observation of a new intermediate redshift damped \lya\
absorption system candidate, discovered in the course of a spectroscopic
follow-up for identifying the sources detected in a 150{\AA}-wide bandpass 
UV-imaging survey at 2000~\AA . The system displays very strong \mgii\ and
\feii\ lines and a high \feii/\mgii\ ratio, which, following photoionization
models, indicates a very high neutral hydrogen column density. Such kind of
systems being very rare at redshifts $\le1.7$, but of prime importance for
understanding the evolution of star formation in galaxies, the newly discovered
candidate deserves further investigations in a near future.
\end{abstract}   
%
\section{\label{intro}Introduction}
	The damped \lya\ Absorption Systems (DLAS), detected in the spectra of
high redshift quasars are characteristic of large neutral hydrogen  column
densities, and were first thought to arise in protogalactic disks (Wolfe et al.
1986). This vision  has recently evolved with the detection of low redshift
systems for which optical identification of the absorber could be performed
(Steidel et al. 1994, Le Brun et al. 1997, Rao \& Turnshek 1998). It appears in 
fact that all morphological types of galaxies are likely to give rise  to such
absorptions, and it becomes possible to study in detail the characteristics of
the interstellar medium of these galaxies (absorption line analysis gives
informations on metallicity, ionization level, velocity structure ...), coupled
to their morphological type. Therefore, these DLAS are very powerful tools for
studying the evolution of galaxies in the aspect of gaseous content,
metallicity and  star formation history : the absorbing gas detected in the 
galaxies is at the origin of stars, and the evolution of its amount gives clues
for star formation rate estimates vs. cosmic time (see e.g Lanzetta et al.
1995). 

	However, in the redshift interval $z<1.7$ (which covers up to 77\% of
the age of the Universe), only very few DLAS are known : 1 from the HST ``QSO
Absorption Lines Key Project'' (Jannuzi et al. 1998), 2 from the IUE QSO survey
(Lanzetta et al. 1995), and 7 inferred from the properties of metal lines
and/or 21~cm absorption, and confirmed a posteriori with HST spectroscopy
(Boiss\'e et al. 1998). As  a result of a HST UV spectroscopic survey of strong
\mgii\ systems, Rao \& Turnshek (1998) have increased the sample with 12 new 
absorbers, but there is anyway a clear need for more absorbers to sample
correctly the full properties of the gas vs. galaxy morphological type,
abundance, etc.

	It as recently been shown that the presence of an absorbing galaxy
close to a quasar sightline may induce a bias in the optical
magnitude-limited
samples of quasars (for example, the dust contained in the galaxy may lead to a
substantial extinction of the quasar, Boiss\'e et al. 1998), and it is therefore
important to investigate different ways of selection for the quasars. In this
context, we present our analysis of a UV-selected quasar spectrum displaying
strong and partially resolved absorption lines by the Mg{\sc ii} doublet at a
redshift of $z_{\rm a}\simeq 0.808$. 

        Given the scarcity of so strong absorption features in known QSOs as
mentioned above, we have searched successfully for other metal absorption
lines. In this paper, we briefly present the available data 
in Section~2, while a
preliminary analysis of the properties of the absorption system is
given in Section~3,
leading to the conclusion that it is very likely a new candidate to the
DLAS sample that consequently deserves further spectroscopic and imaging 
observations at higher resolution, and towards more UV wavelengths. We
also emphasize the high efficiency of space UV experiments for
identifying the $z\la 1.5$ DLAS.
\begin{figure*}
\centerline{\psfig{figure=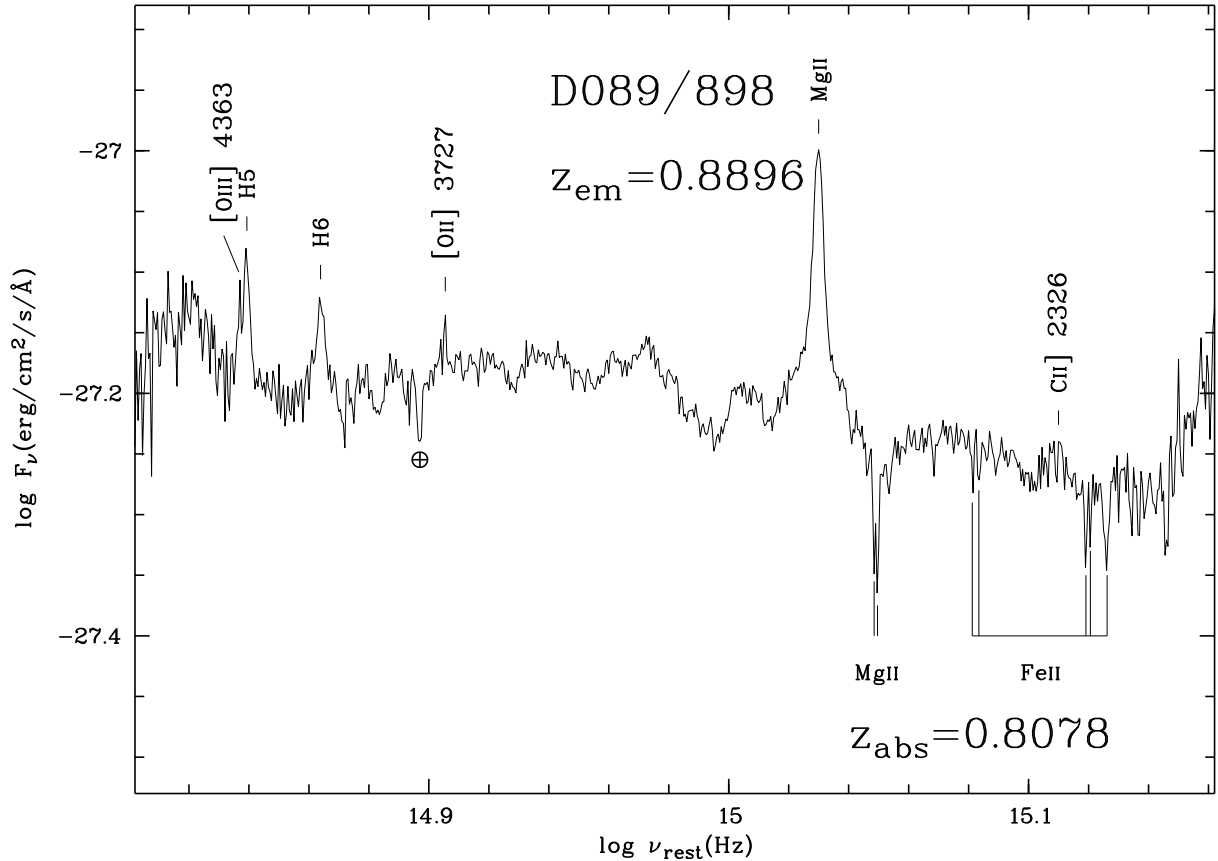,width=18cm,angle=-90,clip=true}}
\caption{\label{spec} Optical spectrum of QSO D089/898. The absolute flux
calibration is based on the POSS2  B$\simeq19.04$ magnitude value. The
absorption lines listed in Table~\ref{lines} are labelled.}
\end{figure*}
\section{Presentation of the data}
	The new DLAS candidate has been identified thanks to the FOCA
balloon-borne survey at 200\,nm (see Milliard, Donas \& Laget, \cite{MDL}).
This 40~cm telescope has been routinely flown over years and has revealed a
large set of UV sources with monochromatic magnitudes $m_{200}\la 18.5$. Most
of them ($\simeq70\%$) are extragalactic, as recently confirmed by a
spectroscopic follow-up (Milliard et al. 1998), performed in the optical domain toward a high galactic
latitude ($+53^\circ$) field of 0.6 square degree including the extragalactic
cluster Abell 2111 (A2111). 

        Optical spectroscopy of the FOCA's UV-detected candidates around A2111
has been carried out with the NORRIS multifiber spectrograph (Hamilton et al.,
1993\cite{DH}) mounted at the Cassegrain f/16 focus of the Hale 5m telescope at
Mount Palomar. A detailed description of the observations from 1995 through 
1997, their calibration and methods of reduction will be given in a forthcoming
paper (Milliard et al. \cite{BM}), but for the sake of clarity of the present
paper let us just mention that the wavelength domain was
$\simeq3800-9000$\,{\AA} with a resolution varying from $\approx7-12$\,{\AA}
FWHM over the full CCD field, and normally two exposures of $50-60$\,min were
taken on the sky, each one being bracketed by 2 FeAr arc exposures for
wavelength calibration, while flat-fielding was achieved using dome-flats.

        Special attention was payed to the sky background subtraction, so
critical in the infrared, and a relative flux calibration was provided by the
very hot sdO standard BD$+28^{\circ}4211$, while the continuous and selective
atmospheric absorptions have been approximately corrected for, with
satisfactory results in terms of spectral flux distribution over the full
wavelength domain in most cases, and in particular for D089/898 as can be
checked on Fig.~1.

	Of the 113 field extragalactic objects spectroscopically observed
around A2111,  8 are QSOs (previously unidentified) with redshifts in the range
$0.8\la z \la1.5$. One of them is FOCA's id. number D089/898, a $B\simeq19$
magnitude object that shows a strong, unresolved emission of the Mg{\sc ii}
2796-2803\,{\AA} doublet at $z_{\rm e}\simeq0.890$ (together with several
weaker emission lines, see Fig.~1), but also partially resolved absorptions by
the Mg{\sc ii} doublet at a lower redshift of $z_{\rm a}\simeq 0.808$, and with
rest equivalent widths $\ga1$\,{\AA}. 

	Notice that the only deep optical image presently available for this
field is a B-band POSS2 image. With a limiting magnitude of 22, it is not deep
enough neither for identifying the absorbing galaxy nor constraining the
brigthness. 
\section{Characteristics of the system}
\begin{table*}
\caption{\label{lines} Absorption lines detected}
\smallskip
\begin{tabular}{lllll}
\hline\noalign{\smallskip}
$\lambda_\mathrm{obs}$(\AA) & $\Wo$(\AA)$^\mathrm{a}$ & Identification & $z$ & $\Wr$ \\
\hline\noalign{\smallskip}
 4236.7$^\mathrm{b}$ & (0.7-0.9) & Fe{\sc ii} 2343 & & \\
 4291.48  &      0.72 & Fe{\sc ii} 2374 & 0.8079  & $0.4\pm 0.2$ \\
 4306.44  &      1.26 & Fe{\sc ii} 2382 & 0.8079  & $0.7\pm 0.2$ \\
 4676.57  & (0.6-0.8) & Fe{\sc ii} 2586 & 0.8085  & \\
 4699.38  & (0.5-1.2) & Fe{\sc ii} 2599 & 0.8070  & \\
 5053.71  &      2.43 & Mg{\sc ii} 2796 & 0.80784 & $1.3 \pm 0.2$ \\
 5067.00  &      2.05 & Mg{\sc ii} 2803 & 0.80789 & $1.1 \pm 0.2$ \\
\noalign{\medskip}\hline
\end{tabular}\par\noindent
$^\mathrm{a}$ All $W_{\rm obs}$ values measured at positions calculated using
the mean redshift of 0.80787 adopted for the Mg{\sc ii} doublet\\
$^\mathrm{b}$ in unknown blend or anomalous noise trough
\end{table*}

	The absorption lines detected in the quasar spectrum are presented in
Table~1. Besides the strong Mg{\sc ii} doublet, several lines of Fe{\sc ii} are
the only other conspicuous features at our spectral resolution and $S/N$ ratio.
Upper limits on the strength of the Mg{\sc i} 2852\,{\AA} (0.6\,{\AA}) and
Ca{\sc ii} 3934\,{\AA} (0.3\,{\AA}) lines have also been estimated. 

	The strengths of the \mgii\ and \feii\ lines strongly suggest the 
system is a DLAS. Photoionization models (Bergeron \& Stasi\'nska 1986), show
that the presence of \feii\ lines is a good indicator of high neutral hydrogen
column densities, and that whenever the \feii/\mgii\ ratio (measured by 
$\Wr(\mbox{\feii\,2382})/\Wr(\mbox{\mgii\,2796})$) is greater than about unity,
the absorber is very likely a strong DLAS ($\nhi \ga 2\,10^{20}$\cm2), where a \feii/\mgii\ ratio
greater than about 0.6 indicates a neutral hydrogen column density above $10^{20}$\cm2
(Bergeron, private communication). The criterion  has been successfully used
by Wampler et al. (1993), and by Le Brun et al. (1997) to study the lower
redshift DLAS, for which the \lya\ line was in the UV range, and thus
unobserved. In the $z_{\rm a}=0.808$ absorber present in the spectrum of
D089/898, we measure $\mbox{\feii/\mgii} \simeq 0.52$, a value comparable to
the lower limit inferred by Bergeron \& Stasi\'nska (1986) for the DLAS. 

	We have also listed in Table~\ref{absorbers} the rest equivalent widths
and the \feii/\mgii\ ratio of the already known intermediate redshift DLAS,
together with the candidate DLAS recently detected at $z=0.558$ in the spectrum
of PKS~0118-272 by Vladilo et al. (1997). As can be seen, the system discovered
in the spectrum of D089/898 has line rest equivalent widths and a \feii/\mgii\
ratio within the range of the other systems. All the confirmed systems have
$\nhi \ge 2\, 10^{20}$\cm2, and the candidate DLAS in the spectrum of
PKS~0118$-$272 displays Ca\,{\sc ii} and Ti\,{\sc ii} lines, which are only
detected in DLAS. Taken all together, these facts indicate that the system 
detected in our UV-selected QSO spectrum is very likely a DLAS.

\begin{table*}
\caption{\label{absorbers} \mgii\ and \feii\ absorption
lines in low redshift systems}
\smallskip
\begin{tabular}{lllllllllll}
\hline\noalign{\smallskip}
Quasar & $\zd$ &\multicolumn{2}{c}{\mgii}& \mgi & \multicolumn{5}{c}{\feii}&
\feii/\mgii \\
 && 2796 &  2804 &  2853 &  2344 &  2374 & 
2382 &  2586 &  2600\\
\hline\noalign{\smallskip}
EX 0302-223 & 1.0095 & 0.88 & 0.86 & 0.17 & $-$  & $-$  & $-$  & 0.57 & 0.62 &
0.70$^\mathrm{a}$\\
PKS 0454+039& 0.8596 & 1.08 & 1.03 & $-$  & 0.94 & 0.72 & 1.32 & 1.31 & 1.10 & 1.22 \\
PKS 1229-021& 0.3950 & 2.22 & 1.93 & 0.43 & $-$  & $-$  & $-$  & 0.64 & 1.50 & 0.68$^\mathrm{a}$ \\
3C 286      & 0.8490 & 0.65 & 0.51 & 0.44 & 0.17 & 0.29 & 0.38 & 0.47 & 0.36 & 0.58 \\
Q 1209+107  & 0.6295 & 2.91 & 2.05 & $-$  & 1.60 & 0.70 & 1.93 & 1.09 & 1.50 & 0.66 \\
3C 196      & 0.4370 & 2.00 & 1.88 & 0.88 & 1.48 & 1.08 & 1.57 & 1.68 & 1.75 & 0.78 \\
PKS 0118$-$272& 0.558& 0.49 & 0.46 & 0.16 & 0.27 & 0.16 & 0.34 & 0.26 & 0.36 & 0.70 \\
D089/898    & 0.8078 & 1.34 & 1.13 & $-$  & $-$  & 0.40 & 0.70 & $-$  & $-$  & 0.52 \\
\hline\noalign{\smallskip}
\end{tabular}
\par\noindent
$^\mathrm{a}$ \feii/\mgii \ ratio estimated from the \feii\,2600 line
\end{table*}

	Because of the small size of the FOCA/NORRIS QSO sample, it is
legitimate to address the questions of the probability to find such DLAS 
candidates, and of favorable effects possibly at play in our selection process.
As a comparison, the HST/FOS Key Project did discover only one system among 35
quasars. But half of them have redshifts below 0.7 (of which 6 have 
$z\le 0.3$), and thus the HST redshift path is very short. In
contrast, the actual FOCA sample of QSOs around A2111 contains only 8
objects but with redshifts in the range $[0.8,1.5]$. Given that i) the NORRIS
spectroscopic observations allow to observe \mgii\ absorption lines with $z>
0.42$, yielding a cumulative redshift path in the FOCA sample of 5.45; and ii)
the average number of DLAS at $\left<z\right>=0.64$ being $\left<n(z)\right> =
0.083\pm0.046$ (Lanzetta et al. 1995), one expects to detect $0.63\pm 0.35$
DLAS, a value compatible with our detection of one system. Indeed, it must be
emphasized that compared to the HST/FOS Key Project, we are probing higher
redshifts that give access to larger $z$ paths in regions of higher DLAS
density. 

	More generally, ground-based spectroscopic follow-up of VUV imaging
surveys is an efficient way to discover DLAS candidates with
measurable \lya\ lines up to $z \simeq 1.7$ , since the detection in a VUV 
bandpass picks up with a
high probability QSOs that are bright enough in the region of  the 
\lya\ absorption line, including the scarce high redshift ones that are 
unaffected by the intergalactic Lyman continuum opacity 
(M\o ller \& Jakobsen 1990). In this context, the GALEX
project (planned for launch in 2001 as a NASA Small Mission Explorer) will
provide a VUV survey quite suited to such a program in its imaging mode. In
addition, with a spectroscopic mode that can by itself discover $z<1.5$ DLAS from
their \lya\ line directly measured in the UV data, GALEX is expected to increase
the present sample of known DLAS by an order of magnitude.  

\acknowledgements{MV and BM wish to thank Chris Martin (PI on GALEX) for the opportunity
of observing with NORRIS, together with the whole team at Palomar, and
especially Todd Small (from Cambridge, UK) for his constant and efficient
assistance during the observations. }

\end{document}